\documentstyle[12pt,aaspp4,flushrt]{article}

\def\gtorder{\mathrel{\raise.3ex\hbox{$>$}\mkern-14mu
             \lower0.6ex\hbox{$\sim$}}}
\def\ltorder{\mathrel{\raise.3ex\hbox{$<$}\mkern-14mu
             \lower0.6ex\hbox{$\sim$}}}

\begin{document}
\title{LARGE SCALE STRUCTURE OF THE UNIVERSE}

\author{NETA A. BAHCALL
\\Princeton University Observatory\\Princeton, NJ}



\section*{Abstract}

How is the universe organized on large scales?  How did this structure
evolve from the unknown initial conditions of a rather smooth early
universe to the present time?  The answers to these questions will
shed light on the cosmology we live in, the amount, composition and
distribution of matter in the universe, the initial spectrum of
density fluctuations that gave rise to this structure, and the
formation and evolution of galaxies, clusters of galaxies, and larger
scale structures.

To address these fundamental questions, large and accurate sky surveys
are needed---in various wavelengths and to various depths.  In this
presentation I review current observational studies of large scale
structure, present the constraints these observations place on
cosmological models and on the amount of dark matter in the universe,
and highlight some of the main unsolved problems in the field of
large-scale structure that could be solved over the next decade with
the aid of current and future surveys.  I briefly discuss some of
these surveys, including the Sloan Digital Sky Survey that will
provide a complete imaging and spectroscopic survey of the
high-latitude northern sky, with redshifts for the brightest $\sim
10^6$ galaxies, $10^5$ quasars, and $10^{3.5}$ rich clusters of
galaxies.  The potentialities of the SDSS survey, as well as of
cross-wavelength surveys, for resolving some of the unsolved problems
in large-scale structure and cosmology are discussed.

\section{Introduction}
\label{sec:one}

Studies of the large-scale structure of the universe over the last
decade, led by observations of the
distribution of galaxies and of clusters of galaxies, have revealed  
spectacular results, greatly increasing our
understanding of this subject. With
 major surveys currently underway, the next decade will provide  new
milestones in
 the study of large-scale structure.  I will
highlight
 what we currently know about large-scale structure, emphasizing some
of the
 unsolved problems and what we can hope to learn in the next
 ten years from new sky surveys.

Why study large-scale structure? In addition to revealing 
the ``skeleton'' of our universe, detailed
 knowledge of the large-scale structure provides constraints on  
 the formation and evolution of galaxies and larger structures, and on the
 cosmological model of our universe (including the mass density of the
universe, the
 nature and amount of the dark matter, and the initial spectrum of fluctuations
that
 gave rise to the structure seen today).

What have we learned so
 far, and what are the main unsolved problems in the field of
 large-scale structure?  I discuss these questions in the sections that follow.
I first list some of the most interesting unsolved problems on which
 progress is likely to be made in the next decade using upcoming sky surveys.

\begin{trivlist}
\item[$\bullet$] Quantify the measures of large-scale structure.  
How large are the largest coherent structures?  How strong is the clustering
on
 large scales  (e.g., as quantified by the power spectrum and the correlation
functions of
 galaxies and other systems)?
\item[$\bullet$] What is the topology of large-scale structure?  What are the 
shapes and morphologies of superclusters, voids,
filaments, and their
 networks?
\item[$\bullet$] How does large-scale structure 
depend on galaxy type, luminosity, surface
 brightness? How does the large-scale distribution of galaxies differ
from that of other systems (e.g., clusters, quasars)?
\item[$\bullet$] What is the amplitude of the peculiar velocity field as a
function of scale? 
\item[$\bullet$] What is the amount of mass and
 the distribution of mass on large scales?
\item[$\bullet$] Does mass trace light on large scales?  What is in the ``voids"?
\item[$\bullet$]What are the main properties of clusters of galaxies:
their mass, mass-function, temperature-function, and dynamical state?
\item[$\bullet$] What is the mass density, $\Omega_m \equiv \rho_m/
\rho_{\mathrm{crit}}$, of the
 universe?
\item[$\bullet$] How does the large-scale structure evolve with time?
\item[$\bullet$] What
 are the implications of the observed large-scale structure for
 the cosmological model of our universe and
 for structure formation? (e.g., What is the nature of the dark
matter?  Does structure form by gravitational instability?  What is
the initial spectrum of fluctuations that gave rise to the structure
we see today?  Were the fluctuations Gaussian?)
\end{trivlist}

\section{Clustering and Large-Scale Structure}
\label{sec:two}

\begin{figure}[b]
\vglue.4in
\plotfiddle{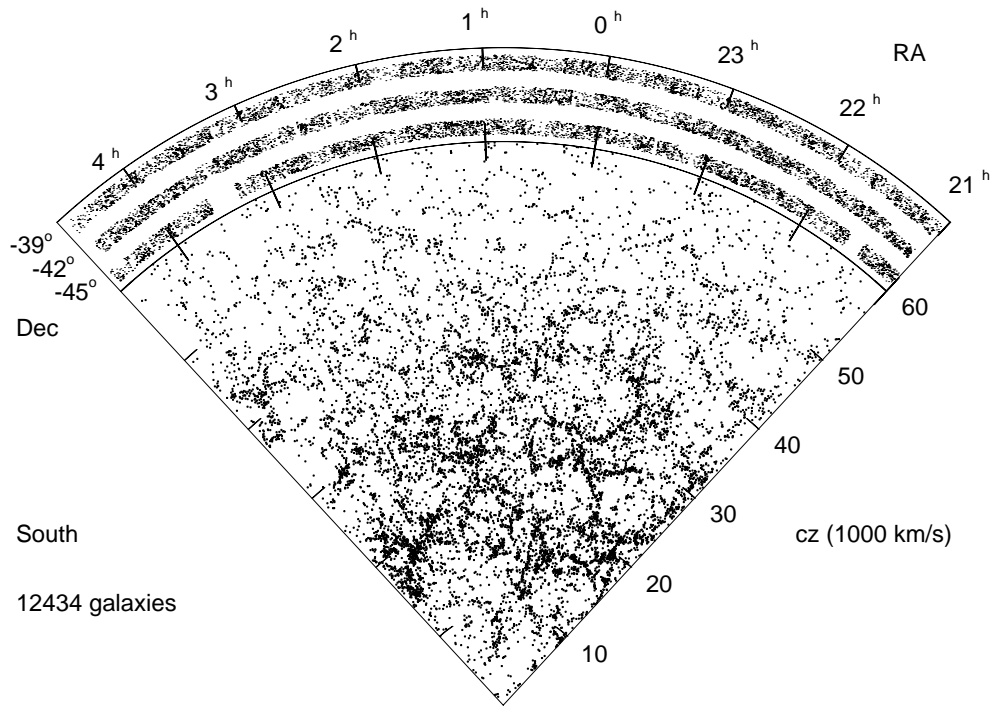}{12cm}{0}{60}{60}{-185}{-90}
\vglue.3in
\caption[]{Redshift cone diagram for 
galaxies in the Las Campanas survey
\cite{kirshner95}\label{fig:one}.}
\end{figure}

Two-dimensional surveys of the universe
 analyzed with 
correlation function statistics \cite{groth77,maddox90} reveal 
structure to
 scales of at least $\sim 20 h^{-1}$~Mpc.  Large redshift surveys
of the
 galaxy distribution reveal a considerably more detailed structure
of superclusters, voids, and filament
 network extending to scales of $\sim 50$--$100h^{-1}$~Mpc
 (\cite{gregory78}--
\nocite{gregory81,chin81,giov86,delap86,dacosta88}\cite{geller89}).  The
 most recent and largest redshift survey, the Las Campanas Redshift
Survey  (\cite{kirshner95}; 
see also \cite{landy96}), with redshifts for $\sim 25 \times 10^3$ galaxies, is
presented in Figure~\ref{fig:one}; it reveals the ``cellular" nature of the 
large-scale galaxy distribution.
The
 upcoming Sloan Digital Sky Survey (SDSS), expected to begin operation
in 1997 
 (see \S~\ref{sec:five}), will provide a three dimensional map of the entire high-latitude
 northern sky to $z \sim 0.2$, with redshifts for approximately $10^6$
galaxies. 
 This survey, and others currently planned, 
will provide the large increase in the
survey volume
  required to resolve some of the unsolved problems
listed above. (See contribution by McKay, this volume.)  

\begin{figure}[t]
\vglue-2.3in
\plotfiddle{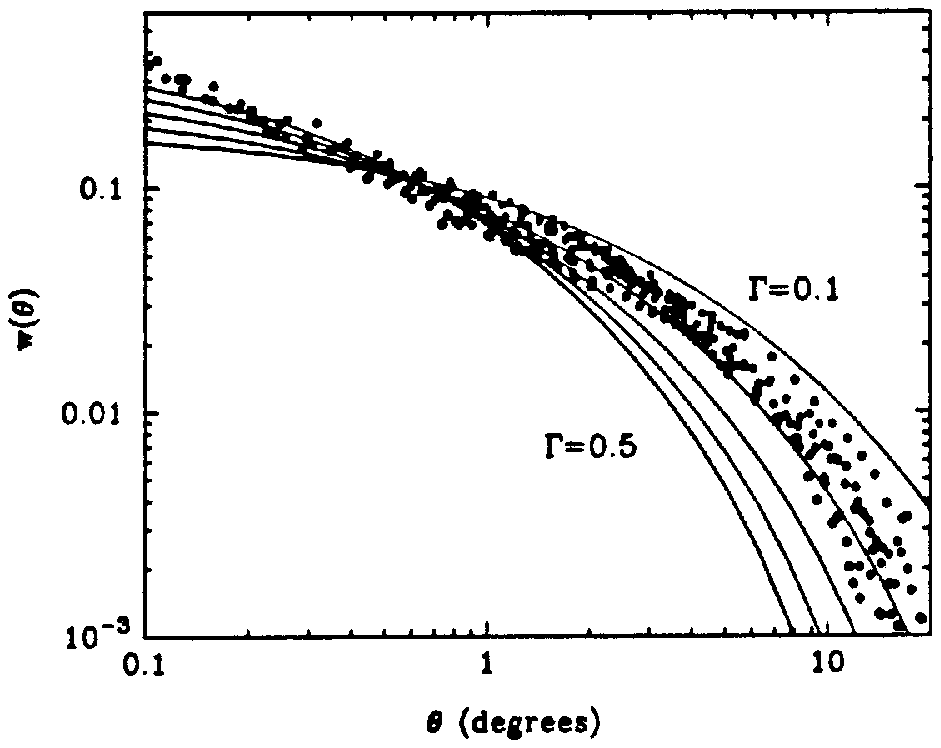}{12cm}{0}{60}{60}{-185}{-90}
\vglue1.5in
\caption[]{The
 scaled angular correlation function of galaxies measured from the APM
survey plotted
 against linear theory predictions for CDM models (normalized to
$\sigma_8 = 1$ on $8 h^{-1}~{\mathrm{Mpc}}$ scale) with
$\Gamma \equiv \Omega_m h = 0.5, 0.4, 0.3, 0.2$ and 0.1
\cite{efs90}\protect\label{fig:two}.}
\end{figure}

The angular galaxy correlation function
 was first determined from the 2D Lick survey and inverted into a
 spatial correlation function by Groth \& Peebles \cite{groth77}.  They
find $\xi_{gg}(r) \simeq 20 r^{-1.8}$ for $r \ltorder 15 h^{-1}$~Mpc, with correlations that
 drop to the level of the noise for larger scales.  This observation implies that
galaxies
 are clustered on at least  $\ltorder 15 h^{-1}$~Mpc scale, with a 
 correlation scale of $r_o(gg) \simeq 5 h^{-1}$~Mpc, where $\xi (r) 
\equiv (r/r_o)^{-1.8} \equiv Ar^{-1.8}$.  More recent results 
support the above conclusions, but show a weak correlation tail to larger scales. The
 recent two-point angular galaxy correlation
function from the
 APM 2D galaxy survey \cite{maddox90,efs90} is presented in 
Figure~\ref{fig:two}.  The
 observed correlation function is compared with expectations from the
cold-dark-matter (CDM) cosmology (using
 linear theory estimates) for different values 
of the 
parameter $\Gamma =\Omega_m h$.  Here 
$\Omega_m$ is the mass density of the universe in
terms of the critical density and  $h \equiv H_0/100~
\mathrm{km~ s^{-1}~
Mpc^{-1}}$.
The different $\Omega_m h$ models differ mainly in
 the large-scale tail of the galaxy correlations:  higher values of
$\Omega_m h$
 predict less structure on large scales (for a given normalization of
the initial mass fluctuation spectrum) since the CDM fluctuation
spectrum peaks
 on scales that are inversely proportional to $\Omega_m h$.  It is clear from
 Figure~\ref{fig:two}, as was first shown from the analysis of galaxy 
clusters (see below),
that
 the standard CDM model with $\Omega_m = 1$ and $h = 0.5$ does
 not produce enough large-scale power to match the observations.
As Figure~\ref{fig:two} shows, the galaxy correlation function
requires $\Omega_m h \sim 0.15$--0.2 for a CDM-type spectrum, consistent with
 other large-scale structure observations.

\begin{figure}[t]
\vglue-1.1in
\plotfiddle{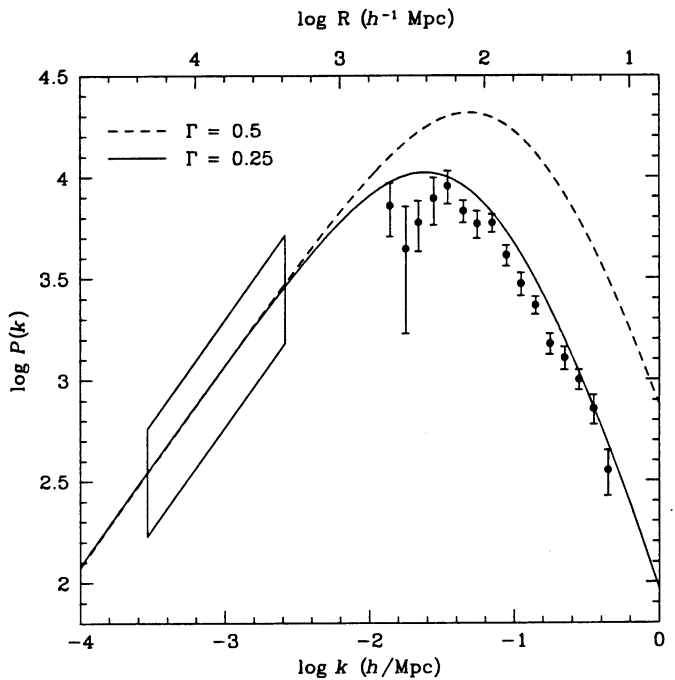}{12cm}{0}{60}{60}{-185}{-90}
\vglue1.5in
\caption[]{The power spectrum as derived from a variety of
 tracers and redshift surveys, after correction for non-linear effects, redshift
distortions, and relative
 biases; from \cite{peacock94}.  The two curves show the
Standard CDM
 power spectrum ($\Gamma = 0.5$), and that of CDM with $\Gamma = 0.25$.  Both are
 normalized to the COBE fluctuations, shown as the box on the
left-hand side of the figure\label{fig:three}.}
\end{figure}

The power spectrum, $P(k)$, which reflects the initial spectrum
of fluctuations that gave rise to
 galaxies and other structure, is
represented by the Fourier transform of the
 correlation function.  One of the recent attempts to determine 
this fundamental statistic using a variety of tracers is presented 
in Figure~\ref{fig:three} (\cite{peacock94}; see
also \cite{landy96},\cite{vogeley92}--\nocite{fisher93}\cite{park94}).  The determination of this composite
spectrum assumes different
 normalizations for the different tracers used (optical galaxies, IR
galaxies, clusters of
 galaxies).  The different normalizations imply a different bias
parameter $b$ for each of the different tracers [where $b \equiv
(\Delta \rho/\rho)_{\mathrm{gal}}/(\Delta \rho/\rho)_m$ represents the
overdensity of the galaxy tracer relative to the mass overdensity]. 
Figure~\ref{fig:three} also shows the microwave background radiation
(MBR) anisotropy as
 measured by COBE \cite{smoot92} on the largest scales 
($\sim 1000 h^{-1}$~Mpc) and compares the data with the mass power
spectrum expected for two CDM models:  a
standard
 CDM model with $\Omega_m h = 0.5\  (\Omega_m = 1, h = 0.5$), and
 a low-density CDM model with $\Omega_m h = 0.25$.  The latter model appears
 to provide the best fit to the data, given the normalizations used
 by the authors for the different galaxy tracers. The recent Las
Campanas redshift survey has reported excess power on $\sim
100H^{-1}$~Mpc scale over that expected from a smooth CDM spectrum
\cite{landy96}. This is a most important observation that will need to
be verified by larger surveys.

The next decade will provide critical advances in the
 determination of the power spectrum and correlation function.
The large redshift
 surveys now underway, the Sloan and the 2dF surveys, 
will probe the power spectrum of galaxies
to
 larger scales than currently available and with greater
accuracy.  These surveys
 will bridge the gap between the current optical determinations of 
$P(k)$ of galaxies on scales $\ltorder 100 h^{-1}$~Mpc and the MBR anisotropy on 
scales $\gtorder 10^3 h^{-1}$~Mpc (see McKay, this volume).  
This bridge will cover the 
 critical range of the
 spectrum turnover, which reflects the horizon scale at the time of
matter-radiation equality.  This will enable the
determination of the initial spectrum of fluctuations at recombination
that gave rise to the structure we see today and will shed light on the
cosmological model parameters that may be responsible for that
spectrum (such as $\Omega_m h$  and the nature of the dark matter). 
 In the next decade,  
 $P(k)$ will also be determined from the MBR anisotropy surveys on small 
scales ($\sim 0.1^\circ$
 to  $\sim 5^\circ$),  allowing a most important
overlap in the determination of the galaxy $P(k)$ from redshift
surveys and the mass $P(k)$ from the MBR anisotropy.  
These data will place  constraints
on cosmological parameters including $\Omega (= \Omega_m +
\Omega_\Lambda), \Omega_m,\
\Omega_b, h$, and the nature of the dark matter itself.

Another method that can efficiently  quantify the large-scale structure of the
universe is the correlation function of clusters of galaxies.  Clusters are correlated in space more strongly than 
are individual galaxies, by
an order of
 magnitude, and their correlation extends to considerably larger
scales ($\sim 50 h^{-1}$~Mpc).  The cluster correlation strength 
increases with richness ($\propto$ luminosity or mass) of the
 system from single galaxies to the richest clusters
\cite{bahcall83,bahcall88}.  
  The correlation strength also
increases with
 the mean spatial separation of the clusters
\cite{szalay85,bahcall86}.  
This dependence results in a ``universal"
dimensionless cluster correlation
 function; the cluster dimensionless correlation scale is constant for
all clusters when
 normalized by the mean cluster separation.
  
Empirically, the two general relations that satisfy the correlation 
function of clusters of galaxies, $\xi_i = A_i
r^{-1.8}$, are: $A_i \propto N_i$, and $A_i \simeq (0.4 d_i)^{1.8}$
\cite{bahcall92a}. 
(Here $A_i$ is
 the amplitude of the cluster correlation function, 
$N_i$ is the richness 
of
 the galaxy clusters of type $i$, and $d_i$ is the mean separation of the
 clusters.)  These observed relations have been compared with
expectations from different cosmological models, yielding 
powerful constraints on the models (see below).

\begin{figure}[t]
\vglue-2.0in
\plotfiddle{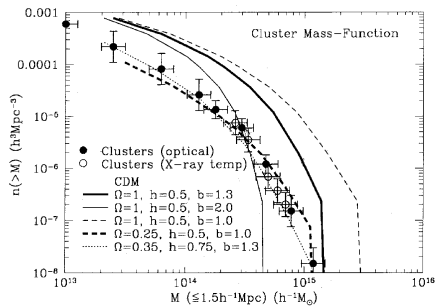}{12cm}{0}{60}{60}{-185}{-90}
\vglue1.5in
\caption[]{The mass function
 of clusters of galaxies from observations (points) and cosmological
simulations of different
 $\Omega_m h$ CDM models \cite{bahcall92b,bahcall93}\label{fig:four}.}
\end{figure}

The observed mass function (MF), $n(>M)$, of clusters of galaxies,
which describes  
the
 number density of clusters above a threshold mass $M$,  
 can also be used as a critical test of theories of structure formation in
the universe.  The richest, most massive clusters are thought to 
form from rare high peaks in the
initial
 mass-density fluctuations; poorer clusters and groups form from
smaller, more common
 fluctuations.  
Bahcall and Cen \cite{bahcall93} determined the MF of clusters 
of galaxies using both optical
and
 X-ray observations of clusters.  
Their MF is presented in Figure~\ref{fig:four}.  
The function is well fit by the analytic expression
\begin{equation} 
n(>M) = 4 \times 10^{-5} (M/M^*)^{-1} \exp (- M/M^*) h^3~
{\mathrm{Mpc^{-3}}}\ ,
\label{eq:four}
\end{equation} 
with $M^* = (1.8 \pm 0.3) \times 10^{14} h^{-1}~ M_\odot$, (where the mass
 $M$ represents the cluster mass within $1.5 h^{-1}$~Mpc radius).

Bahcall and Cen \cite{bahcall92b} compared 
the observed mass function and correlation function
 of galaxy clusters with predictions of N-body cosmological simulations
of
 standard $(\Omega_m = 1)$ and
 nonstandard $(\Omega_m < 1)$ CDM models. They find 
that none of the standard $\Omega_m = 1$~CDM 
 models, with any normalization, can reproduce both the observed 
correlation function and the mass
function of clusters.  A
low-density
 ($\Omega_m \sim 0.2$--0.3) CDM-type model, however, provides a good
 fit to both sets of observations (see, e.g., Fig.~\ref{fig:four}).

\section{Peculiar Motions on Large Scales}
\label{sec:three}

 How is the mass distributed in the universe?  Does it follow, on the
average,  the light
distribution? To address this important question, peculiar motions on large
scales are studied
 in order to directly trace the mass distribution.  It
 is believed that the peculiar motions (motions relative to a pure
Hubble
 expansion) are caused by the growth of cosmic
structures due to gravity.  
A comparison
 of the mass-density distribution, as reconstructed from peculiar
velocity data, with
 the light distribution (i.e., galaxies) provides information on how
well
 the mass traces light \cite{dekel94,strauss95}. 
 A formal 
analysis yields a measure of the
 parameter $\beta \equiv \Omega_m^{0.6}/b$. 
Other   methods that place
constraints on $\beta$
 include the anisotropy in the galaxy distribution in the redshift
direction due to peculiar motions (see \cite{strauss95} for a
review).

Measuring peculiar motions is difficult.  The motions are
usually inferred with the aid of measured distances to galaxies or clusters
that are obtained  using some (moderately-reliable) 
distance-indicators (such as
 the Tully-Fisher or $D_n -\sigma$ relations), and 
the measured galaxy redshift.  The peculiar velocity $v_p$ is 
 then determined from the difference between the measured redshift
velocity, $cz$, and
 the measured Hubble velocity, $v_H$, of the system (the latter obtained from the
distance-indicator):  $v_p = cz - v_H$.

The dispersion
 in the current measurements of $\beta$ is very large. 
No strong conclusion can therefore be reached at present
 regarding the values of $\beta$ or $\Omega_m$.  The larger and more accurate
 surveys currently underway, including high precision velocity
measurements, may lead to the
 determination of $\beta$ and possibly its decomposition into
$\Omega_m$ and $b$ (e.g., \cite{cole94}).

Clusters of galaxies can also serve as
 efficient tracers of the large-scale peculiar velocity field in the
universe \cite{bahcall94b}.  Measurements of cluster peculiar velocities are
 likely to be more
 accurate than measurements of individual 
galaxies, since cluster distances can be
determined by averaging
 a large number of cluster members as well as by using different
distance indicators.  Using large-scale cosmological simulations, 
Bahcall et al.~\cite{bahcall94b}
find that clusters
 move reasonably fast in all the cosmological models studied, tracing well the
underlying matter
 velocity field on large scales. A comparison of model expectation
 with the available data of cluster velocities is presented by Bahcall
 and Oh \cite{bahcall96}.  The current data suggest consistency with
 low-density CDM models.  Larger velocity surveys are needed to
 provide more robust comparisons with the models.

\section{\textsc{Dark Matter and Baryons in Clusters of Galaxies}}
\label{sec:four}

Optical and X-ray observations of rich clusters of
 galaxies yield cluster masses that range from $\sim 10^{14}$ to $\sim 10^{15} h^{-1}
 M_\odot$ within $1.5 h^{-1}$~Mpc radius of the cluster center.  When
 normalized by the cluster luminosity, a median value of $M/L_B \simeq
300 h$ 
is observed for rich clusters.  This mass-to-light ratio implies
 a dynamical mass density of $\Omega_{\mathrm{dyn}} \sim 0.2$ on $\sim 1.5
h^{-1} {\mathrm{~Mpc}}$ scale.  If, as suggested by
 theoretical prejudice, the universe has critical density ($\Omega_m = 1$),
then most of
 the mass in the universe {\it cannot} be concentrated in clusters, groups
 and galaxies; the mass  would have to be distributed 
more
 diffusely than the light.

\begin{figure}[b]
\vglue-2.0in
\plotfiddle{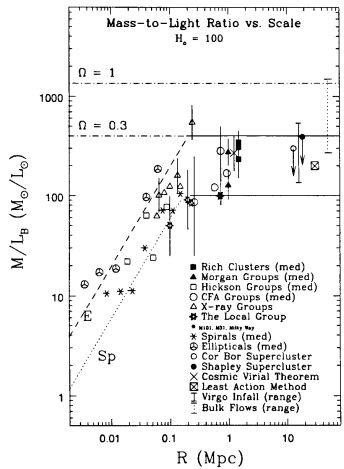}{12cm}{0}{60}{60}{-185}{-90}
\vglue1.5in
\caption[]{A composite mass-to-light ratio 
of different systems---galaxies, groups, clusters, and
superclusters---as a
 function of scale.  See Bahcall et al. 1995 \protect\cite{bahcall95b}
for details.\label{fig:five}.}
\end{figure}

A recent analysis of the mass-to-light
 ratio of galaxies, groups and clusters  \cite{bahcall95b} 
suggests that while the $M/L$ ratio of galaxies increases with scale
 up to radii of $R \sim 0.1$--$0.2 h^{-1}$~Mpc, due
 to the large dark halos around galaxies, this ratio
 appears to flatten and remain approximately constant for groups and
rich clusters, to scales of $\sim 1.5$~Mpc, and possibly even to the 
larger scales of
superclusters (Fig.~\ref{fig:five}).  The flattening occurs at $M/L_B \simeq
200$--$300 h$, corresponding
to
 $\Omega_m \sim 0.2$.  This observation may suggest that most of the dark
 matter is associated with the dark halos of galaxies and that
 clusters do {\it not} contain a substantial amount of additional
dark matter, other
 than that associated with (or torn-off from) the galaxy halos, and
 the hot intracluster medium.  Unless the distribution 
 of matter is very different from the distribution of light, with
 large amounts of dark matter in the ``voids" or on very large
 scales, the cluster observations suggest that the mass density in the universe may be
 low, $\Omega_m \sim 0.2{\rm -}0.3$.

Clusters of galaxies contain many baryons.  
Within $1.5 h^{-1}$~Mpc of a rich cluster, the X-ray 
emitting gas contributes $\sim$~3--$10 h^{-1.5}\%$ of
 the cluster virial mass (or $\sim 10$--30\% for $h=1/2$)
\cite{briel92,white95}.  
Visible stars contribute only a small additional amount to this
 value.  Standard Big-Bang nucleosynthesis limits the mean baryon
density of the
 universe to $\Omega_b \sim 0.015 h^{-2}$ \cite{walker91}.  
This suggests that the baryon fraction
 in some rich clusters exceeds that of an $\Omega_m = 1$ universe by
 a large factor \cite{white93,lubin95}.  Detailed
 hydrodynamic simulations \cite{white93,lubin95} suggest that baryons are
not preferentially segregated into
 rich clusters.  It is therefore suggested that either the
mean density
 of the universe is considerably smaller, by a factor of $\sim 3$, than
 the critical density, or that the baryon density of the universe is
 much larger than predicted by nucleosynthesis.  The observed baryonic
mass fraction in
 rich clusters, when combined with the
 nucleosynthesis limit, suggests   
$\Omega_m \sim 0.2$--0.3; this estimate is consistent with
 the dynamical estimate determined above. Future optical and X-ray sky surveys of
clusters of galaxies should help resolve these most interesting problems. 

\section{The Sloan Digital Sky Survey}
\label{sec:five}

A detailed description of the upcoming Sloan Digital Sky Survey (SDSS)
is presented in these proceedings by McKay.  I will not repeat it
here.  I only summarize that the SDSS is a complete photometric and
spectroscopic survey of $\pi$ steradians of the northern sky, using
30 $2048^2$ pixel CCDs in five colors (${\mathrm{u^\prime,\  
g^\prime,\  r^\prime,\  i^\prime,\ z^\prime}}$), and two spectrographs
($R=2000$) with 640 total fibers.  The 5-color imaging survey will
result in a complete sample of $\sim 5 \times 10^7$ galaxies to a
limiting magnitude of $r^\prime \sim 23^m$, and the redshift survey
will produce a complete sample of $\sim 10^6$ galaxy redshifts to $r^\prime \sim
18^m$ ($z \sim 0.2$), $\sim 10^5$ galaxy redshifts to
$r^\prime \sim 19.5^m$ ($z \sim 0.4$) for the reddest brightest
galaxies, $\sim 10^5$ quasar redshifts to $g^\prime \sim 20^m$, and
$\sim 10^{3.5}$ rich clusters of galaxies.

What are some of the most interesting scientific problems in
large-scale structure that the large and accurate Sloan sky survey can
address?  

\begin{trivlist}
\item[$\bullet$]Quantify the clustering (of galaxies, clusters of
galaxies, quasars) on large scales using various statistics (power
spectrum, correlation function, void-probability distribution, and
more).
\item[$\bullet$]Quantify the morphology of large-scale structure (the
supercluster, void, filament network).
\item[$\bullet$]Determine the distortion in the redshift space
distribution and its implication for the mass-density of the universe.
\item[$\bullet$]Determine the clustering as a function of luminosity,
galaxy type, surface brightness, and system type (galaxies, clusters, quasars).
\item[$\bullet$]Determine the clustering properties of clusters
(superclustering, correlation function and its richness dependence,
power spectrum).
\item[$\bullet$]Study the dynamics of clusters of galaxies. (With the
availability of up to hundreds of redshifts per cluster, the mass of
clusters can be well determined and compared with X-ray and lensing
masses.  The cluster mass-function and velocity function will be
accurately determined, as well as the $M/L$ and
$\Omega_{\mathrm{dyn}}$ implications).
\item[$\bullet$]Study the evolution of galaxies, clusters, and
superclusters to $z \sim 0.5$, and the evolution of quasars to $z
\gtorder 5$.  These should provide important new constraints on cosmology.
\item[$\bullet$]Use all the above to place strong constraints on the
cosmological model and $\Omega$, as discussed in the previous
sections.
\end{trivlist}

\section{Important Future Surveys}
\label{sec:six}

What are some of the important surveys needed in order to address the
main unsolved problems listed in section \ref{sec:one}?  I list below
such surveys.

\begin{trivlist}
\item[$\bullet$]Optical, infrared, and radio redshift surveys (of
galaxies, clusters, quasars, AGNs).  These will help solve the
quantitative description of large-scale structure, its strength and
topology, and the relation among the structures described by different
objects.
\item[$\bullet$]X-ray surveys of clusters, quasars, and 
possibly superclusters.  These will allow a good determination of the contribution of
the hot gas component in the universe, cluster masses and temperature
function, baryon fraction in clusters (and superclusters?), and the
evolution of clusters and quasars.
\item[$\bullet$]Gravitational lensing surveys.  These will allow the
most direct determination of the total mass and mass-density
distribution in galaxies, clusters, and large-scale structure.
\item[$\bullet$]Peculiar motion surveys of galaxies and clusters
should yield most important constraints on $\Omega_m$ and $b$.
\item[$\bullet$]High redshift surveys, using optical ground based
telescopes (Keck), HST, X-rays, and radio, should reveal the important
but yet unknown time evolution of structure in the universe.  This
will provide a fundamental clue to models of galaxy formation and cosmology.
\item[$\bullet$]MBR anisotropy surveys, currently underway, will
provide the fluctuation spectrum of the microwave background radiation and
hopefully determine many of the cosmological parameters such as
$\Omega$, $\Omega_m$, $\Omega_b$, $H_o$, and the initial spectrum of 
fluctuations.
\item[$\bullet$]All the above surveys will greatly constrain, and
possibly determine the cosmological parameters of the universe ($H_o;
\Omega;\ \Omega_b;\ q_o;\ \lambda$).
\end{trivlist}

Research support by NSF grant 93-15368 and NASA grant NGT-51295 is
gratefully acknowledged.

\end{document}